\documentclass[conference]{IEEEtran}

\usepackage{amsmath,amssymb,amsfonts}
\usepackage{algorithmic}
\usepackage{algorithm}
\usepackage{graphicx}
\usepackage{textcomp}
\usepackage{xcolor}
\usepackage[square,numbers,sort&compress]{natbib}
\usepackage{booktabs}
\usepackage{multirow}
\usepackage{hyperref}
\hypersetup{
}
\usepackage{subcaption}
\usepackage{listings}
\usepackage{tabularx}
\usepackage{comment}
\usepackage{amssymb}
\usepackage{xcolor}
\usepackage{fancyhdr}

\usepackage{tikz}
\usetikzlibrary{positioning, arrows.meta, shapes.multipart, fit, backgrounds, calc}

\usepackage{siunitx}

\lstdefinelanguage{SPARQL}{
  morekeywords={PREFIX, SELECT, WHERE, FILTER, DISTINCT, COUNT, AS},
  sensitive=false,
  morecomment=[l]{\#},
  morestring=[b]"
}

\begin{document}
\fancypagestyle{firstpage}{
    \fancyhf{}
    \fancyfoot[C]{\footnotesize\itshape
Accepted author manuscript at the IEEE Annual Conference of the Industrial Electronics Society (IEEE IECON 2026). This version has been accepted for publication and may differ slightly from the final published version.}
    \renewcommand{\headrulewidth}{0pt}
    \renewcommand{\footrulewidth}{0pt}
}



\title{Towards Cyber-Physical Cognition: A Unified Ontology-Driven Knowledge Graph for Real-Time Autonomous Grid Operations}

\author{
  \IEEEauthorblockN{
    Sathvik Sankaranarayanan\IEEEauthorrefmark{1}, 
    Michael Mandulak\IEEEauthorrefmark{2}, 
    Ibrahim Shahbaz\IEEEauthorrefmark{1}, 
    and Eman Hammad\IEEEauthorrefmark{1}\IEEEauthorrefmark{2}
  }
  \IEEEauthorblockA{
    Texas A\&M University, College Station, TX, USA \\
    \IEEEauthorrefmark{1}iSTAR Laboratory, Department of Engineering Technology and Industrial Distribution \\ \IEEEauthorrefmark{2}SPARTA Laboratory, Institute of Data Science
  }
}
\maketitle
\thispagestyle{firstpage}


\begin{abstract}

Modern power systems and smart grids are often composed of fragmented and heterogeneous data silos, which lack the cohesion needed for effective cross-domain analysis. For this, this paper introduces a universal ontology framework for the operational representation of intelligent cyber-physical power systems via a unified knowledge graph and an ontology capable of cross-domain reasoning. This work focuses on bridging cyber-physical simulators as a stepping stone towards that vision. By establishing a unified semantic middleware grounded in IEC 61970 (CIM) and IEC 62351/61850 standards, this framework integrates disparate cyber and physical simulation environments, illustrated via OMNeT++ and PowerWorld, into a single knowledge graph. Evaluation across three standard power system benchmarks demonstrates sub-linear scaling in both knowledge graph size and construction time. We further validate the framework's efficacy for real-time decision support, achieving millisecond-level query performance across both domains, maintained across six cumulative structural mutations to the knowledge graph. The resulting unified knowledge graph provides a robust, scalable information corpus for autonomous smart grid operations, enabling complex analysis of real-world power systems.

\end{abstract}


\section{Introduction and Motivation}
\label{sec:intro}

The increasing penetration of distributed energy resources, electrified transportation, and intelligent control systems has made modern power systems far more dynamic, interconnected, and data-intensive than before~\cite{al2026engineering}. Traditional power system models used in studies and digital twins are based primarily on numerical simulation and static network representations, which often struggle to capture the complex interactions among physical, cyber, and human components within cyber-physical systems (CPS). Coupled with the ever-growing scale and heterogeneity of data in dispatch and operation, manual or purely algorithmic approaches are increasingly inadequate for real-time analysis and decision support in smart grids ~\cite{chen2022,Liu2023A}.

The current challenges are characterized by the following three factors: \textit{(1) Data fragmentation}, where simulation outputs, operational logs, and sensor data remain isolated across disparate domains and temporal resolutions, complicating end-to-end process modeling; \textit{(2) Lack of semantic interoperability}, in which simulation environments fail to integrate heterogeneous models across electrical, communication, and market domains, often resulting in ontologies that only address isolated subsystems; and \textit{(3) Limited temporal representation}, where static system models are unable to capture continuously evolving behaviors or contextual changes within a time series.

\begin{figure*}
    \centering
    \includegraphics[width=0.9\linewidth]{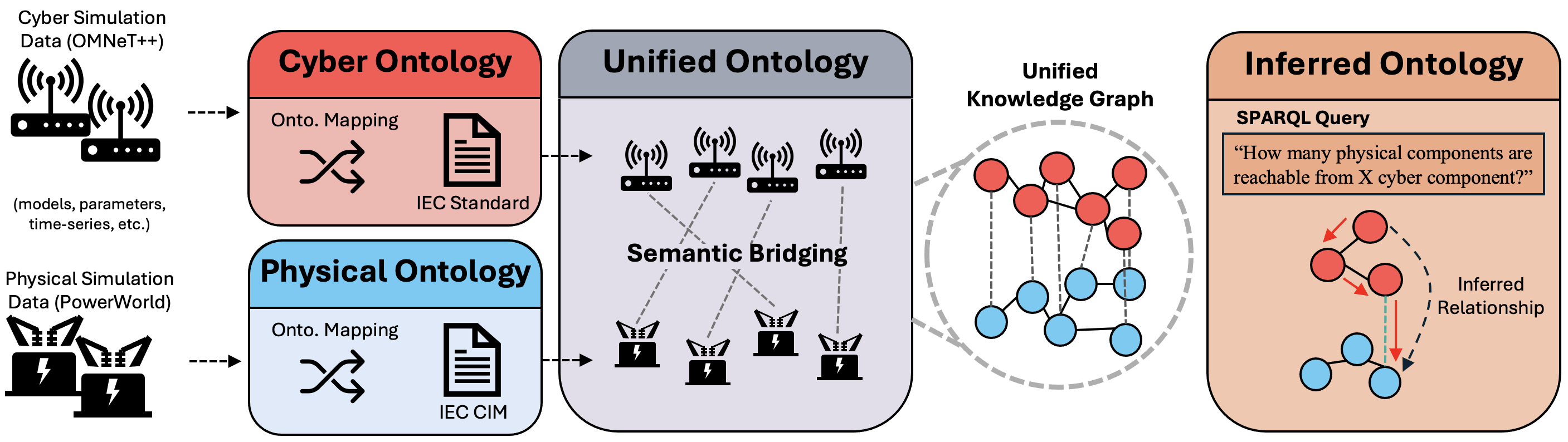}
    \caption{Proposed framework for physical and cyber ontology alignment for a unified onotology and knowledge graph Representation. The cyber and physical ontologies'smapping of real-time simulator data to the CIM, is  aligned into the unified ontology and queries and validation refinements produce a final structure representation as a knowledge graph.}
    \label{fig:workflow}
    \vspace{-7pt}
\end{figure*}

To address these challenges, this paper proposes an ontology-driven knowledge graph (KG) approach, integrating semantic modeling with real-time cyber and physical simulation data for power system modeling. Visually outlined in Fig. \ref{fig:workflow}, this unified framework consolidates information from diverse sources—power and communication simulators, documentation, and sensor data—into a scalable, extensible representation grounded in the Common Information Model (CIM). Through the bridging of these domains, the proposed ontology framework will enable applications in real-time decision reasoning, predictive analysis, optimization, and anomaly detection in complex cyber-physical power systems. 

Our contributions are summarized as follows: 1) 
\begin{itemize}
    \item \textit{Cyber-Physical Unified Ontology:} Development of a unified schema that maps heterogeneous simulation outputs to CIM-standardized classes.
    \item \textit{Benchmark Statistics for CPS Ontologies:} Demonstrated sub-linear end-to-end runtime scalability for ontology generation and sub-linear scaling with system size.
    \item \textit{Efficient and Resilient Querying:} Real-time querying of cyber, physical and cross-domain relationships in the order of milliseconds, validated under sequential structural updates to the knowledge graph.
\end{itemize}



\section{Background and Related Works}

At the intersection of power systems, CPS and data modeling, popular methods map singular domain data instances to structured data schemas, some of which are discussed here.

\textit{Ontologies for Power Systems:} Ontologies act to unify data formats by providing a formal, machine-readable specification of data and the relationships between them. Typically built off of the Web Ontology Language (OWL) semantic web standard, these ontologies focus on unifying domain data for logical reasoning. Specifically, power system ontologies aim to model complex system dynamics, processes and operations in CPS to enable real-time analytics and decision support within the system domain. Towards this, the Common Information Model (CIM) IEC 61970/61968 \cite{iec61970_301, iec61968_11} provides a standardized vocabulary and schema to facilitate data modeling across heterogeneous power systems. The IEC 61970/61968 is recognized as the Smart Grid ontology, providing a standardized UML-based semantic model of power system objects and their relationships  \cite{Wang2011An,Schumilin2017Towards,Gomez2017CIM}. 
The CIM has been furthered in numerous works to accurately capture increasingly complex power grid data while others have developed notable frameworks such as Grid2Onto \cite{grid2onto} and OntoPowSys \cite{ontopowsys}, which focus on physical power system components and smart grid communication, respectively. While effective, these frameworks each model a single domain 
with no cross-layer identities between electrical and communication entities. Concretely, Grid2Onto contains no cyber layer, so a query such as 'which physical bus is monitored by a high-risk RTU' is unanswerable within its schema. OntoPowSys similarly carries no IEC 62351 security attributes, meaning properties such as protocol risk or encryption status have no representation. We address both gaps by considering both layers from simulator data in a singular unified ontology, as demonstrated by competency questions X2 and X3 in \ref{tab:competency},  which we discuss in detail in Section \ref{sec:framework}.

\textit{Power System Knowledge Graphs:}
 Knowledge graphs have developed as an integral component of CPS modeling, given the number of complex dynamics associated with real-world systems. For developing these knowledge graphs, the Resource Description Framework (RDF) is a popular method for knowledge graph construction, representing data as relational triples (Subject, Predicate, Object) \cite{Gibbins2017Resource,Tomaszuk2020RDF} for graph-based modeling. In power system applications, RDF-based knowledge graphs capture topologies, component metadata and system interactions into a schema that supports queries and analytics for grid decision support \cite{Kor2023Integrating}. This has also been extended to applications in digital twins \cite{Ramonell2023Knowledge}, cybersecurity \cite{Peng2024Research} and energy CPS \cite{Aryan2021Explainable}, among many others.

Despite these advancements, current research lacks a universal, simulator-agnostic middleware capable of unifying heterogeneous cyber-physical data into a single, high-fidelity knowledge graph. Existing ontologies often operate as disconnected domain silos, focusing on synchronization rather than providing the deep semantic interoperability required for autonomous reasoning. Furthermore, traditional models prioritize static structural representations, failing to incorporate the real-time security metadata and low-latency query performance essential for agentic AI-in-the-loop applications. This paper addresses these gaps by proposing a scalable, standards-compliant framework that bridges disparate simulation environments, such as PowerWorld and OMNeT++~\cite{hammad2019implementation}, into a unified, cognitive information corpus grounded in IEC 61970 and 61850 standards.

\section{Ontology Framework Overview}
\label{sec:framework}

\begin{figure*}[t]
\centering
\includegraphics[width=0.95\linewidth]{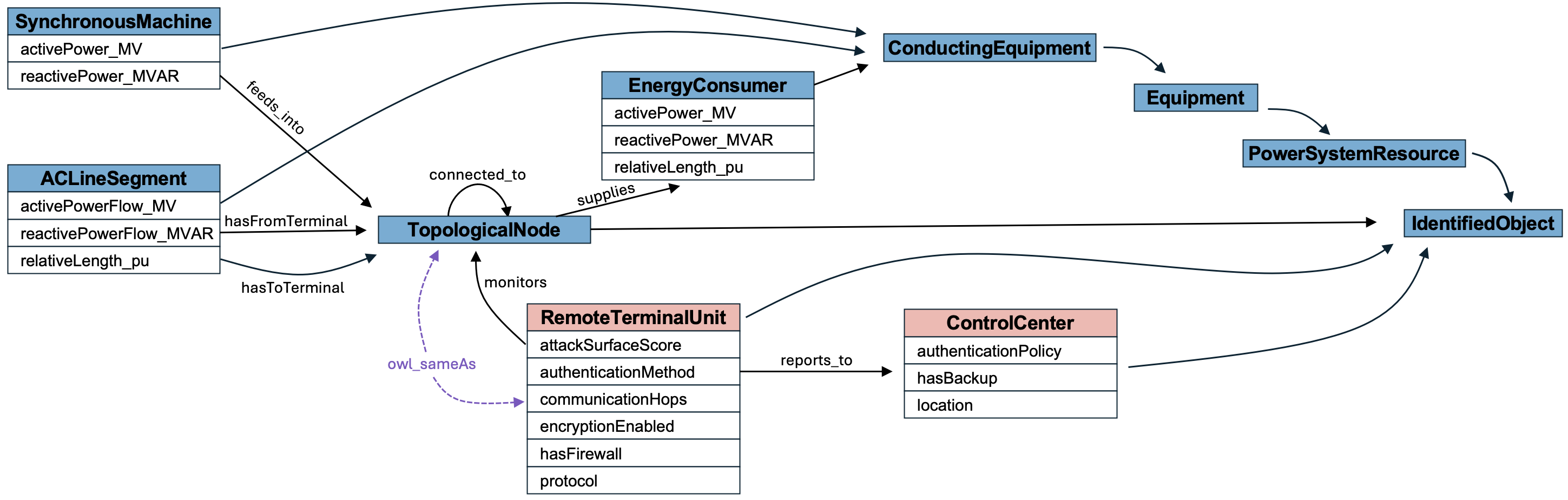}
\caption{Unified cyber-physical ontology rendered directly from the IEEE~39 TTL file.}
\label{fig:tbox}
\end{figure*}

The framework is structured as four layers, each serialized as a separate artifact over one shared TBox. Layers are kept separate rather than fused into a single graph so that each remains independently validatable against its own standard, and so downstream consumers can select the level of integration they require without paying reasoning cost for domains they do not query. The shared TBox is what makes this tractable: cyber classes are rooted at cim:IdentifiedObject rather than in an independent hierarchy, so cyber and physical entities are schema-compatible by construction and require no post-hoc schema alignment. Throughout, we retain the standard separation of Terminological (TBox) statements, which fix the standards-compliant schema, from Assertional (ABox) statements, which instantiate a specific solved system state. As illustrated in Fig.~\ref{fig:tbox}, the layers are:

\begin{itemize}
\item \textit{Physical Layer:} Grounded in the IEC 61970 Common Information Model (CIM) \cite{iec61970_301}, capturing the physical electrical components.
\item \textit{Cyber/SCADA Layer:} Grounded in IEC 62351 and IEC 61850 \cite{iec61850_7_4}, representing communication infrastructure, protocols, and security metadata.
\item \textit{Unified Layer:} Merges cyber and physical layers via semantic bridging to induce cross-layer connections.
\item \textit{Inferred Layer:} Introduces formal logic and reasoning to extract non-trivial relationships, creating a consistent, actionable information corpus.
\end{itemize}

A knowledge graph is chosen over flat relational schemas because power system queries inherently span entity types — a single contingency analysis may join buses, lines, RTUs, and security scores — and RDF's triple structure natively supports that without schema migrations. The formal TBox/ABox split then allows each domain layer to be validated independently against its own standard while sharing a common schema backbone.
This structure ensures that the resulting knowledge graph (KG) can accurately represent the system components relative to the CIM and IEC standards. Each layer is detailed in the following subsections.


\subsection{Physical Ontology}
\label{sec:phys-onto}



Formally, the physical ontology is the pair 
$\mathcal{O}_{\text{phys}} = (\mathcal{T}_{\text{phys}}, \mathcal{A}_{\text{phys}})$, 
where the schema/vocabulary is defined by the tuple 
$\mathcal{T}_{\text{phys}} = \langle \mathcal{C}_{p}, \mathcal{R}_{O,p}, \mathcal{R}_{D,p} \rangle$ 
with $\mathcal{C}_{p}$ being the set of CIM classes, $\mathcal{R}_{O,p}$ the object properties, 
and $\mathcal{R}_{D,p}$ the data properties. 

Let $\mathcal{I}_{p}$ be the set of named individuals (representing buses, lines, 
machines, loads, and nodes) and $\mathcal{V}$ 
be the set of XSD (XML Schema Definition) literals. The ABox $\mathcal{A}_{\text{phys}}$ is defined 
as a finite set of ground assertions such that:
\[
\begin{aligned}
  \mathcal{A}_{\text{phys}} \subseteq {} & \{ C(a) \mid C \in \mathcal{C}_{p}, \, a \in \mathcal{I}_{p} \} \\
  & \cup \{ r(a,b) \mid r \in \mathcal{R}_{O,p}, \, a, b \in \mathcal{I}_{p} \} \\
  & \cup \{ d(a,v) \mid d \in \mathcal{R}_{D,p}, \, a \in \mathcal{I}_{p}, \, v \in \mathcal{V} \}.
\end{aligned}
\]

Through this definition, we can represent any topology of physical system components through a set of CIM classes $\mathcal{C}_p$, XML literals $\mathcal{V}$ and relationships between the objects $\mathcal{R}_{O,p}$ and structure of the data $\mathcal{R}_{D,p}$. To populate the CIM classes, we define a mapping from PowerWorld's Easy SimAuto (ESA) data to the CIM for standardization. This mapping is shown in brief in Table \ref{tab:esa-cim}.

\begin{table}[h]
\centering
\caption{Alignment of PowerWorld Easy SimAuto (ESA) data to the Common Information Model (CIM). CIM classes carry the \texttt{cim:}
prefix; data-property suffixes (\texttt{\_pu}, \texttt{\_kV},
\texttt{\_deg}, \texttt{\_MW}, \texttt{\_MVAR}) are omitted for conciseness.
All literals are \texttt{xsd:float}.}
\label{tab:esa-cim}
\small
\renewcommand{\arraystretch}{1.1} 
\newcommand{\tree}{\color{gray}\llcorner\,} 

\begin{tabular}{ll}
\toprule
\textbf{Easy SimAuto (ESA)} & \textbf{Common Information Model (CIM)} \\
\midrule
\texttt{bus} & \texttt{TopologicalNode} \\
\quad $\tree$ \texttt{PUVolt} & \quad $\tree$ \texttt{voltageMagnitude} \\
\quad $\tree$ \texttt{Angle} & \quad $\tree$ \texttt{voltageAngle} \\
\quad $\tree$ \texttt{kV} & \quad $\tree$ \texttt{nominalVoltage} \\
\midrule
\texttt{branch} & \texttt{ACLineSegment} \\
\quad $\tree$ \texttt{LineMW} & \quad $\tree$ \texttt{activePowerFlow} \\
\quad $\tree$ \texttt{LineMVR} & \quad $\tree$ \texttt{reactivePowerFlow} \\
\quad $\tree$ \texttt{LineLength} & \quad $\tree$ \texttt{relativeLength} \\
\midrule
\texttt{gen} & \texttt{SynchronousMachine} \\
\quad $\tree$ \texttt{GenMW} & \quad $\tree$ \texttt{activePower} \\
\quad $\tree$ \texttt{GenMVR} & \quad $\tree$ \texttt{reactivePower} \\
\midrule
\texttt{load} & \texttt{EnergyConsumer} \\
\quad $\tree$ \texttt{LoadMW} & \quad $\tree$ \texttt{activePower} \\
\quad $\tree$ \texttt{LoadMVR} & \quad $\tree$ \texttt{reactivePower} \\
\bottomrule
\end{tabular}
\vspace{-15pt}
\end{table}

\subsection{Cyber Ontology}

The cyber ontology is formally defined in a similar manner to that of the physical ontology:
$\mathcal{O}_{\text{cyber}} = (\mathcal{T}_{\text{cyber}}, \mathcal{A}_{\text{cyber}})$. 
The cyber schema/vocabulary is defined by the tuple 
$\mathcal{T}_{\text{cyber}} = \langle \mathcal{C}_{\text{c}}, \mathcal{R}_{O,\text{c}}, \mathcal{R}_{D,\text{c}} \rangle$, 
where $\mathcal{C}_{\text{c}}$ is the set of cyber-infrastructure classes (e.g., routers, 
switches, RTUs, communication links, and measurement channels), $\mathcal{R}_{O,\text{c}}$ 
is the set of cyber object properties, and $\mathcal{R}_{D,\text{c}}$ is the set of cyber data properties.

Let $\mathcal{I}_{\text{c}}$ be the set of named cyber individuals extracted from the 
network configuration file, and let $\mathcal{V}$ remain the set of XSD literals. The cyber 
ABox $\mathcal{A}_{\text{cyber}}$ is defined as a finite set of ground assertions such that:
\[
\begin{aligned}
  \mathcal{A}_{\text{cyber}} \subseteq {} & \{ C(c) \mid C \in \mathcal{C}_{\text{c}}, \, c \in \mathcal{I}_{\text{c}} \} \\
  & \cup \{ r(c_1, c_2) \mid r \in \mathcal{R}_{O,\text{c}}, \, c_1, c_2 \in \mathcal{I}_{\text{c}} \} \\
  & \cup \{ d(c, v) \mid d \in \mathcal{R}_{D,\text{c}}, \, c \in \mathcal{I}_{\text{c}}, \, v \in \mathcal{V} \}.
\end{aligned}
\]

In a similar manner to $\mathcal{A}_{phys}$, $\mathcal{A}_{cyber}$ represents the same relationships and structure of data, but on arbitrary cyber components $c_1, c_2$, while referencing the IEC 62351 and 61850 standards instead of the CIM. We design a mapping to accurately capture data from OMNeT++ in the IEC standards, which we refer to Table \ref{tab:omnet-cim} for a brief description.

  \begin{table}[h]
  \centering
  \caption{Alignment of OMNeT++ network topology data to the IEC 62351 \& 61850 cyber standards.
  All length-derived values are \texttt{xsd:float}; identifiers are
  \texttt{xsd:string}.}
  \label{tab:omnet-cim}
  \small
  \renewcommand{\arraystretch}{1.1}
  \newcommand{\tree}{\color{gray}\llcorner\,}

  \begin{tabular}{ll}
  \toprule
  \textbf{OMNeT++} & \textbf{IEC Standard 62351} \\
  \midrule
  \texttt{node} & \texttt{RemoteTerminalUnit} \\
  \quad $\tree$ \texttt{node\_id} & \quad $\tree$ \texttt{mRID} \\
  \quad $\tree$ \texttt{ip} (auto) & \quad $\tree$ \texttt{ipAddress} \\
  \midrule
  \texttt{link} & \texttt{CommunicationLink} \\
  \quad $\tree$ \texttt{node\_a} & \quad $\tree$ \texttt{fromDevice} \\
  \quad $\tree$ \texttt{node\_b} & \quad $\tree$ \texttt{toDevice} \\
  \quad $\tree$ \texttt{len\_km} & \quad $\tree$ \texttt{latency\_ms} \\
  \bottomrule
  \end{tabular}
  \vspace{-10pt}
  \end{table}

\subsection{Unified Ontology}
\label{sec:unified-onto}

\begin{figure}[t]
\centering
\includegraphics[width=0.95\columnwidth]{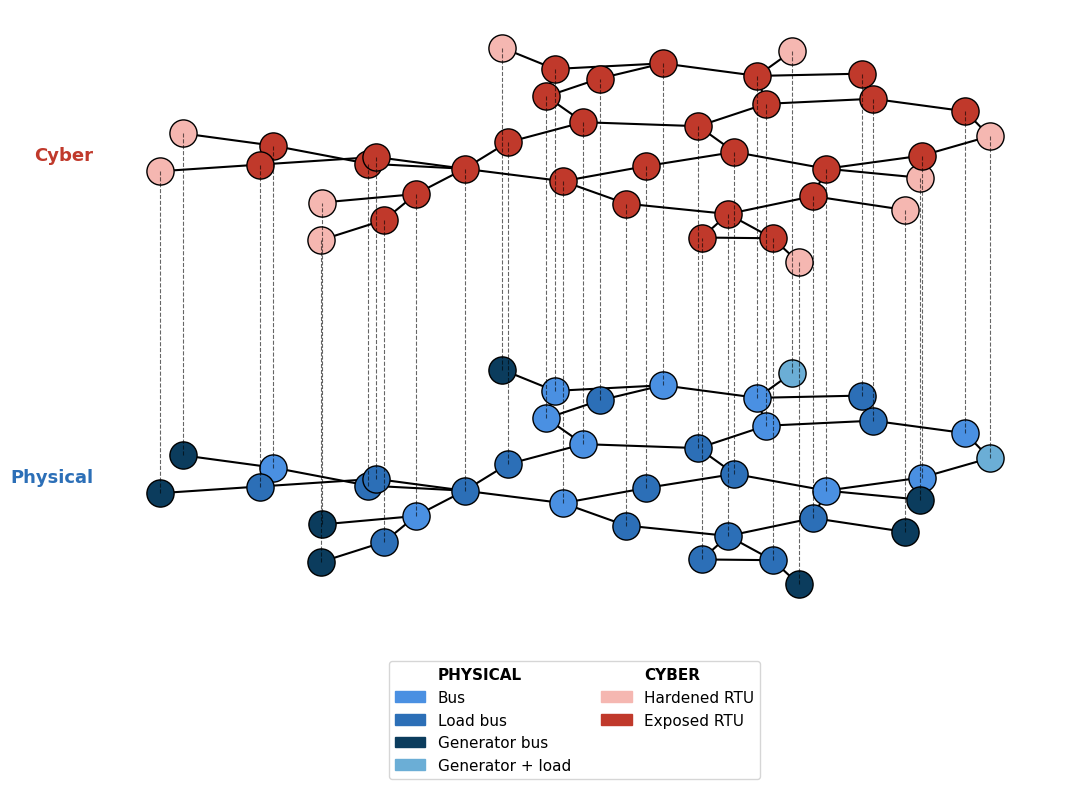}
\caption{Unified cyber-physical graph for the IEEE 39-bus system. Physical buses are linked to RTUs via \texttt{owl:sameAs} bridges. The \textit{Physical} component types and the \textit{Cyber} component risk levels are distinguished by shade.}
\label{fig:kg}
\vspace{-15pt}
\end{figure}

To unify the physical and cyber ontologies, we define a set of bridges between individual physical and cyber components as a one-to-one matching. Formally, this unified ontology is defined as:
\[
\mathcal{O}_{\text{uni}} = \mathcal{O}_{\text{phys}} \cup
\mathcal{O}_{\text{cyber}} \cup \mathcal{B}
\]
where $\mathcal{B}$ represents the bridges. We draw these bridges by aligning unique identifiers in physical and cyber components (Internationalized Resource Idenitifers) with local names and populating a reference for the respective physical and cyber data on the opposite side. The resulting mapping is applied uniformly across each ontology, generating our unified ontology. We provide an example on the IEEE 39-bus system in Figure \ref{fig:kg}, where buses are linked to RTUs in association with cyber risk scores.

\subsection{Inferred Ontology}
\label{sec:inferred-onto}

The unified graph as constructed is intentionally split, keeping physical and cyber components in their respective ontology with unifying bridges between the two. The inferred ontology aims to enable deeper reasoning by merging data into singular entities representing both physical and cyber properties. For this, we run the HermiT reasoner \cite{HermiT}, an ontology reasoner for knowledge inference and alignment, on the unified ontology. This merges the physical and cyber layers while including data relationships between the layers and the individual components, allowing for topology-wide reasoning. We expand on the validation and use cases in Section \ref{sec:results}.

\section{Evaluations}
\label{sec:results}
This section evaluates the proposed physical, cyber and resultant unified ontology. We first describe our experimental platform and data inputs, followed by ontology validation and benchmark statistics for our framework. Results shown were collected primarily using the unified ontology, though we note that the results are consistent across ontologies. 

The ontology framework was developed using Python 3.12. Ontology construction and serialization are performed using \texttt{rdflib} 7.4.0, while logical reasoning, consistency checking and inference are handled using \texttt{owlready} 0.50 with the bundled HermiT reasoner (v.1.4.3.456) \cite{HermiT}. SHACL (Shapes Constraint Language) validation is performed using the \texttt{pyshacl} 0.31.0 engine with SPARQL-based constraint definitions \cite{w3c-shacl}. Ontology visualizations (Fig.~\ref{fig:tbox}) are rendered directly from the unified TTL using Graphviz, an open-source graph layout toolkit, ensuring the figure stays consistent with the underlying ontology. 

\begin{table}[h!]
\centering
\caption{Benchmark system characteristics.}
\label{tab:benchmarks}
\begin{tabular}{lccc}
\toprule
\textbf{Case} & \textbf{Buses} & \textbf{Branches} & \textbf{Generators} \\
\midrule
Kundur & 10 & 9 & 4 \\
IEEE 39  & 39 & 46 & 10 \\
IEEE 118 & 118 & 179 & 54 \\
\bottomrule
\end{tabular}
\vspace{-7pt}
\end{table}

\textit{Data:} We evaluate the scalability and structural integrity of the proposed framework using three standard power system benchmarks: the Kundur Two-Area system \cite{kundur1994power}, the IEEE 39-bus 'New England' system \cite{IEEE39}, and the IEEE 118-bus system \cite{IEEE118bus}. These benchmarks are selected to represent a spectrum of grid complexities, ranging from small-scale stability studies (Kundur) to large-scale, multi-area transmission network topologies (IEEE 118). The physical and cyber characteristics of each test case are summarized in Table \ref{tab:benchmarks}.

\subsection{Ontology Validation}
To validate the proposed ontologies, we verify information retrieval and structural alignment through three methods: \textit{(1)} domain-grounded competency questions, \textit{(2)} model structural validators and \textit{(3)} tests under dynamic conditions. 

\begin{table}[ht] 
\centering
\caption{Competency Questions for Ontologies}
\label{tab:competency}
\scriptsize 
\setlength{\tabcolsep}{3pt} 
\renewcommand{\arraystretch}{1.3}
\begin{tabular}{|p{0.5cm}|l|p{6cm}|} 
\hline
\textbf{ID} & \textbf{Category} & \textbf{Competency Question} \\ \hline

P1 & \multirow{6}{*}{Physical} & How many buses host at least one generator? \\ \cline{1-1} \cline{3-3}
P2 &                           & How many branches (lines + transformers as ACLineSegment) exist? \\ \cline{1-1} \cline{3-3}
P3 &                           & What distinct nominal voltage levels exist in the system? \\ \cline{1-1} \cline{3-3}
P4 &                           & How many load instances are modeled? \\ \cline{1-1} \cline{3-3}
P5 &                           & What is the total active power generation in the system? \\ \cline{1-1} \cline{3-3}
P6 &                           & How many buses have a nominal voltage assigned? \\ \hline

C1 & \multirow{5}{*}{Cyber}    & How many RTUs are instantiated? \\ \cline{1-1} \cline{3-3}
C2 &                           & How many RTUs are hardened (score $<$ 0.1)? \\ \cline{1-1} \cline{3-3}
C3 &                           & How many RTUs use DNP3 Secure Authentication? \\ \cline{1-1} \cline{3-3}
C4 &                           & What distinct attack-surface tiers does the classifier produce? \\ \cline{1-1} \cline{3-3}
C5 &                           & How many control centers do RTUs report to? \\ \hline

X1 & \multirow{3}{*}{\shortstack[l]{Cross-\\Layer}} & How many physical buses are bridged to cyber counterparts? \\ \cline{1-1} \cline{3-3}
X2 &                              & Which load buses are monitored by weakly-secured RTUs (score $>$ 0.5)? \\ \cline{1-1} \cline{3-3}
X3 &                              & How many generator buses are monitored by DNP3\_SA RTUs? \\ \hline

\end{tabular}
\vspace{-10pt}
\end{table}

\textit{Competency Questions:} To assess the validity of our ontologies, we map power system data to a defined set of competency questions (see Table \ref{tab:competency}), spanning cyber, physical and cross-topology instances. In the \textit{Physical} cases (P1-P6), we refer to the physical system components linked as \texttt{PowerSystemResource}, including generators and buses. The \textit{Cyber} cases (C1-C5) cover the properties of the RTUs for power control, and are mapped as \texttt{RemoteTerminalUnit} units linked to a physical component. The \textit{Cross-Layer} instances (X1-X3) cover queries including both cyber and physical components within the unified ontology. Together, these questions mirror the queries an EMS operator, SCADA security analyst, and AI dispatch agent would each need answered in a real operational scenario. To demonstrate alignment with competency questions, we generate SPARQL queries for each question spanning data from both the cyber and physical system components. We provide a visual example in Listing \ref{lst:cq3} for question C3 in Table \ref{tab:competency}. Our competency questions and respective SPARQL queries are all successfully answered by the proposed unified ontology. A complete listing of SPARQL queries is available with the implementation code.

\begin{lstlisting}[
    caption={SPARQL query for competency question C3}, 
    label={lst:cq3},
    numbers=left,
    xleftmargin=2em,
    numbersep=8pt,      
    framexleftmargin=1.5em, 
    breaklines=true      % Prevents code from bleeding into the next column
]
PREFIX cim:   <http://iec.ch/TC57/CIM#>
PREFIX pkg:   <http://powerdypkg.org/ontology#>
PREFIX cyber: <http://powerdypkg.org/cyber#>
PREFIX owl:   <http://www.w3.org/2002/07/owl#>

SELECT (COUNT(*) AS ?n) WHERE {
    ?bus a cim:TopologicalNode .
    ?bus pkg:supplies ?load .
    ?bus owl:sameAs ?cyber_bus .
    ?rtu cyber:monitors ?cyber_bus .
    ?rtu cyber:attackSurfaceScore ?score .
    FILTER (?score > 0.5)
}
\end{lstlisting}

\textit{Data Alignment:} To extend validation, we test our ontologies using both SHACL and HermiT for data and logic assurance, respectively. For SHACL, we define constraints relative to domain restrictions from both cyber and physical aspects based on the CIM. For example, within the \texttt{Topological Node} representation of physical buses, we enforce association with a voltage, magnitude and angle. This is consistent with the linked cyber components, as each bus must be associated with a valid RTU and an \texttt{attackSurfaceScore}. Between units, we also ensure routing logic by enforcing \texttt{hasFromTerminal} and \texttt{hasToTerminal} origin and destinations along transmission lines. Relative to these constraints, our ontologies showed full alignment across all validation test instances. Similar testing using the HermiT logic validator showed full model fit within the unified ontology.

\begin{table}[htbp]
  \centering
  \caption{Query times following dynamic updates to the knowledge graph for the unified ontology of the IEEE 39-bus system. Reported times are taken as an average across 10 tests for each competency question following each structural update in sequence.}
  \label{tab:kg_update_res}
  \begin{tabular}{c l c S[table-format=2.1] S[table-format=1.1] S[table-format=2.1]}
    \toprule
    & & & \multicolumn{3}{c}{\textbf{Query Time (ms)}} \\
    \cmidrule(lr){4-6}
    \textbf{Step} & \textbf{Dynamic Update} & \textbf{Triples} & {\textbf{P1}} & {\textbf{C1}} & {\textbf{X1}} \\
    \midrule
    0 & Unified IEEE 39-Bus      & 1384  & 13.4 & 5.6 &  4.5 \\
    \midrule
    1 & \texttt{remove bus-5}    & 1342   & 14.2 & 5.8 & 4.0 \\
    2 & \texttt{update phys-attr} & 1342   & 16.7 & 6.4 &  4.5 \\
    3 & \texttt{add bus-40}      & 1364   & 13.0 & 5.9 &  4.5 \\
    4 & \texttt{remove bus-15}   & 1327   & 13.4 & 6.5 &  3.9 \\
    5 & \texttt{update cyber-attr}    & 1327   & 12.6 & 5.7 &  4.4 \\
    6 & \texttt{add bus-41}      & 1349   & 14.4 & 5.3 &  3.7 \\
    \bottomrule
  \end{tabular}
  \vspace{-10pt}
\end{table}

\textit{Dynamic Settings:} In the presence of structural or informational updates, the KG must maintain efficiency to accurately capture real-time CPS dynamics. Thus, we further validate our unified ontology by inducing a sequence of additions (\texttt{add}), removals (\texttt{remove}) and attribute updates (\texttt{update}), testing query times following each. We present the results in Table \ref{tab:kg_update_res}. Notably, the average query times remain consistent amidst updates to the KG structure when compared to the unified baseline. At each dynamic update, we also confirm validity using SHACL and HermiT, which maintains that the ontology is logically valid despite disruptions to the KG's structure.

\subsection{Ontology Statistics}
We also provide our ontology statistics as a benchmark for power systems, as is standard in other domains \cite{HPCOntology,LLMCQs}. These statistics include \textbf{(1)} scale as a relative size measure and \textbf{(2)} execution runtimes in generation and querying. We discuss each as follows.

\begin{table}[h!]
\centering
\caption{Ontology triples per benchmark.}
\label{tab:scale}
\begin{tabular}{lrrr}
\toprule
\textbf{Metric} & \textbf{Kundur} & \textbf{IEEE 39} & \textbf{IEEE 118} \\
\midrule
Physical Triples     & 331 &  951 &    2,913   \\
Cyber Triples        & 133 &  423 &     1,213  \\
Unified Triples        & 445 & 1,384 & 4,215 \\
Inferred Triples     & 607 & 1,988 & 6,186 \\
\bottomrule
\end{tabular}
\vspace{-10pt}
\end{table}

\textit{Scale:} We provide the ontology triple counts in Table \ref{tab:scale} across the tested datasets for the physical, cyber, unified and inferred ontologies. The inferred ontology is derived from the asserted graphs using the HermiT reasoner, which logically deduces implicit relationships and class memberships. Across these test cases, the TBox schema remains consistent while ABox size scales linearly with power system complexity. In the smaller Kundur Two-Area system, we observe a reduced triple count in the unified ontology due to overlap in cyber and physical components. At scale for the larger IEEE 39 and 118 systems, the overlap is balanced out by the additional triples for the coverage of cross-domain information necessary for alignment. However, these cases, only introduce 10 and 89 extra triples for IEEE 39 and 118, respectively, preserving required cross-domain properties while maintaining efficiency.

\begin{table}[h]
\centering
\caption{Power System Ontology Runtime Performance }
\label{tab:build_time}
\small 
\addtolength{\tabcolsep}{-2pt} 
\begin{tabular}{l|ccc|c} 
\toprule
& \multicolumn{4}{c}{\textbf{Time}} \\
\cmidrule(lr){2-5}
\textbf{Case} & \textbf{Build (s)} & \textbf{Valid. (s)} & \textbf{Total (s)} & \textbf{Query (ms)} \\
\midrule
Kundur   & 7.14 & 1.06 & 8.20 & 5.75 \\
IEEE 39  & 6.79 & 1.32 & 8.11 & 6.58 \\
IEEE 118 & 7.88 & 1.91 & 9.79 & 9.07 \\
\bottomrule
\end{tabular}
\vspace{-10pt}
\end{table}

\textit{Execution Time:} We report runtime statistics in Table~\ref{tab:build_time} for the ontology build, validation, and end-to-end times, alongside query times for each power system. Query times are averaged across one question from each category from Table~\ref{tab:competency}: \emph{``How many buses host at least one generator?''} (P1), \emph{``How many RTUs are instantiated?''} (C1), and \emph{``How many physical buses are bridged to cyber counterparts?''} (X1). All times are taken as an average of 3 runs.
Notably, the end-to-end execution remains under 10 seconds across all test instances, demonstrating sub-linear scaling. Build times average approximately 84\% of the total runtime
across test instances. Regarding query times, the ontologies support real-time system decisions, with queries averaging 9--10\,ms. These query times also scale sub-linearly relative to the
number of triples, increasing approximately 58\% from Kundur to IEEE~118.

\section{Conclusion \& Future Directions}

This paper presented a universal, tool-agnostic ontology framework designed to transform fragmented cyber-physical  data into a unified, standards-compliant knowledge graph. As a stepping stone, the work focuses on bridging cyber-physical simulators of power systems. By establishing a semantic middleware grounded in the physical and cyber standards (e.g. \textit{IEC 61970 (CIM)} and \textit{IEC 61850/62351}), we have demonstrated a robust methodology for bridging the gap between raw numerical simulation outputs and interconnected representations. Our evaluation across multiple IEEE benchmarks confirms that this unification achieves \textit{sub-linear scaling} in both construction time and graph size, while maintaining millisecond-level query performance amidst dynamic updates. To support real-world smart grid applications, this framework enables multi-domain analytics across heterogeneous data in cyber-physical power systems. By structuring power system cyber and physical components into a unified knowledge graph, this framework provides a foundation for downstream applications, including graph-based machine learning, dynamic topology updates, and agentic reasoning. The inclusion of dynamics such as timestamped updates, failure models and operational conditions across both cyber and physical components can be represented as a singular entity and conform to reasoning engines stemming from knowledge graphs. 

\begin{small}
\bibliographystyle{IEEEtranN}
\bibliography{refs_min}
\end{small}

\end{document}